\documentclass[aps,pre,twocolumn,superscriptaddress]{revtex4}
\usepackage{epsfig,amssymb,amsmath,amsthm,amsfonts,amsbsy,mathrsfs,bm}
\usepackage{graphicx}
\usepackage{color}
\usepackage{siunitx}
\usepackage{bm}

\begin{document}
\title{Accurate determination of the translational correlation function of two-dimensional solids}
\author{Yan-Wei Li}
\affiliation{Division of Physics and Applied Physics, School of Physical and
Mathematical Sciences, Nanyang Technological University, Singapore 637371,
Singapore}
\author{Massimo Pica Ciamarra}
\email{massimo@ntu.edu.sg}
\affiliation{Division of Physics and Applied Physics, School of Physical and
Mathematical Sciences, Nanyang Technological University, Singapore 637371,
Singapore}
\affiliation{
CNR--SPIN, Dipartimento di Scienze Fisiche,
Universit\`a di Napoli Federico II, I-80126, Napoli, Italy
}
\date{\today}

\begin{abstract}
The identification of the different phases of a two-dimensional (2d) system, which might be in solid, hexatic, and liquid, requires the accurate determination of the correlation function of the translational and of the bond-orientational order parameters.
According to the Kosterlitz-Thouless-Halperin-Nelson-Young (KTHNY) theory, in the solid phase the translational correlation function decays algebraically, as a consequence of the Mermin-Wagner long-wavelength fluctuations.
Recent results have however reported an exponential-like decay.
By revisiting different definitions of the translational correlation function commonly used in the literature, here we clarify that the observed exponential-like decay in the solid phase results from an inaccurate determination of the symmetry axis of the solid; 
the expected power-law behaviour is recovered when the symmetry axis is properly identified. 
We show that, contrary to the common assumption, the symmetry axis of a 2d solid is not fixed by the direction of its global bond-orientational parameter, and introduce an approach allowing to determine the symmetry axis from a real space analysis of the sample.
\end{abstract}

\maketitle
\section{Introduction}
Solids posses both translational and bond-orientational orders.
The translational order evaluates the spatial periodicity of the point pattern identified by the position of the molecules, while the bond-orientational order evaluates variations in the local orientation of the pattern.
Hence, translational order implies bond-orientational one, while the converse is not true. 
While both the translational and the bond-orientational orders are lost as a solid melts into a liquid, their variations across a melting transition have non-universal features.
In particular, in three dimensions the translational and the bond-orientational order parameters generally vary synchronously, while this is not always the case in 2d.
Indeed, in 2d a hexatic phase with short-range translational correlations and quasi-long-range bond-orientational correlations is frequently observed. 
If present, this phase is in between the liquid one, where both order parameters are short-ranged, and the solid one, where the bond-orientational order parameter is long-ranged while the translational order parameter is quasi-long-ranged, as a consequence of the Mermin-Wagner long-wavelength fluctuations~\cite{Mermin}.

According to the celebrated KTHNY theory~\cite{KT, HN, Y}, the solid-hexatic and the hexatic-liquid transitions are both continuous, respectively driven by the unbinding of dislocation pairs, and by the dissociation of dislocations into disclinations. The KTHNY melting scenario has been observed both in  experiments~\cite{Maret1999,Keim2013,Keim2014} and in simulations~\cite{Krauth2015,Glotzer,NingXu,John_Russo,Pablo2019}.
However, melting may also proceed via the so-called mixed scenario, where a continuous solid to hexatic transition is followed by a first-order hexatic to liquid transition. 
This mixed scenario has been observed in hard disks~\cite{Krauth2011}, and later in a number of different systems~\cite{Krauth2015,Tanaka,John_Russo,OurPaper,Glotzer,Experiment_harddisc,NingXu,Hajibabaei2019,Pablo2019,weikaiqi}. 
Furthermore, melting may also occur via a first-order solid-liquid transition without any hexatic phase ~\cite{Tanaka,OurPaper,Glotzer,John_Russo}.
Many properties of a system have been show to influence its melting scenario, including
the softness~\cite{OurPaper} and the range~\cite{Krauth2015} of the interaction, density~\cite{NingXu}, polydispersity~\cite{John_Russo,Pablo2019}, energy dissipation~\cite{Tanaka}, shape and symmetry of particles~\cite{Glotzer}, and so on~\cite{active_melting,Hajibabaei2019,weikaiqi}.

The identification of the melting scenario of a given system critically relies on the ability to differentiate the possible phases, via the investigation of the equation of state and the ordering properties of the system.
To distinguish the solid from the hexatic phase one might in principle rely on the investigation of the correlation function of the bond-orientational order parameter, $c_6(r)$.
Indeed~\cite{KT, HN, Y}, the bond-orientational correlation function has no decay in the solid phase, and decays as $c_6(r) \propto r^{-\eta_6}$ with $0 < \eta_6 \leq 1/4$ in the hexatic one.
Practically, however, this approach inevitably leads to a large error in the identification of the phase boundary, as in finite systems it is difficult to reliably estimate when $\eta_6>0$.
For this reason, it is convenient to rely on the translational correlation function, $c(r)$, as this is predicted~\cite{KT, HN, Y} to decay as a power-law in the solid phase, $c(r) \propto r^{-\eta}$ with $\eta \leq 1/3$, and exponentially in the hexatic one, $c(r) \propto \exp(-r/\xi)$.

The accurate evaluation of the correlation function $c(r)$ is however difficult. First, one needs to investigate large systems, as the decay length $\xi$ could be large. 
In addition, the correlation function $c(r)$ depends on a wavevector, or on a direction in space, which needs to be accurately selected. 
As an example of how delicate is the study of $c(r)$, we notice that recent investigations of the melting of 2d Lennard-Jones (LJ) solids suggested $c(r)$ to decay exponentially even in the solid phase~\cite{Hajibabaei2019}. 
A faster than expected decay of $c(r)$ has also been observed in other systems~\cite{Pablo2019}.

In this manuscript, we revisit and compare different definitions of the correlation function of the translational order parameter recently considered in the literature.
We show that the translational correlation functions decaying faster than expected, in the solid phase, are found when the symmetry axis of the crystal is not accurately determined. We demonstrate that this symmetry axis is not, as commonly assumed, fixed by the direction of the global bond-orientation of the sample.
We introduce a novel approach to determine the symmetry axis and show that, when this novel approach is used, the translational correlation function exhibits the expected power-law decay in the solid phase. 

The paper is organized as follows.
Sec.~\ref{sec:numerics} gives detail on the numerical model we use to demonstrate our findings, 
and on the protocol we use to assure we reach the condition of thermal equilibrium.
Sec.~\ref{sec:wavevector} illustrates that, in the solid phase, the position of the first peak of the static structure factor shifts with respect to that of the hexagonal lattice, and demonstrates that this shift must be taken into account to properly evaluate the correlation function of the translational order parameter.
In Sec.~\ref{sec:orientation}, we consider the validity of a recently introduced simple definition of the correlation function of the translational order parameter, which assumes the sample to be oriented along the direction of the global bond-orientation. We show that this assumption is generally not valid, thus rationalizing contrasting results observed in the literature, and discuss how the sample orientation should be determined from a real space analysis.
Finally, we draw our conclusions and recommendations as concern the evaluation of the correlation function of the translational order parameter in Sec.~\ref{sec:end}.
\begin{figure}[!t]
 \centering
 \includegraphics[angle=0,width=0.45\textwidth]{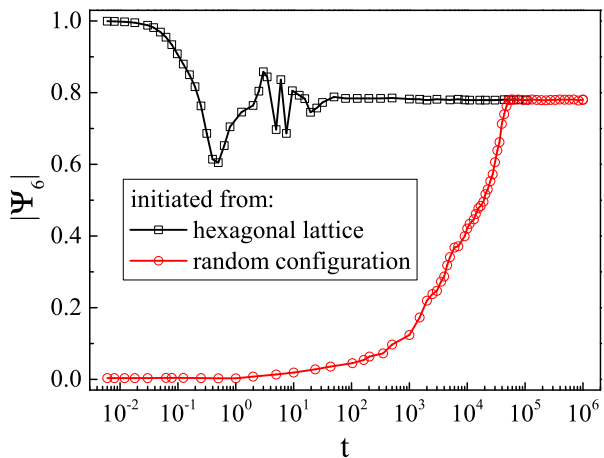}
 \caption{Time evolution of the magnitude of the global order parameter $|{\bf \Psi}_{6}|$ in simulations initiated from a hexagonal lattice configuration (black squares) and a random configuration (red circles). $|{\bf \Psi}_{6}|$ converges at $0.78$ at long time scales. The data is for $N=512^{2}$.
\label{fig:Equ}
}
\end{figure}

\section{Simulation details~\label{sec:numerics}}
We study the solid phase of monodisperse LJ particles of mass $m$, interacting with potential
\begin{equation}
U(r)=
\begin{cases}
4\epsilon[(\sigma/r)^{12}-(\sigma/r)^{6}+C]& r\leq r_c \\
0& \text{otherwise},
\end{cases}
\label{eq:lj}
\end{equation}
where $r_c = 2.5\sigma$, $C$ is a constant chosen such that $U(r_c)=0$.
$\sigma$, $m$ and $\sqrt{m\sigma^{2}/\epsilon}$ will be our units of length, mass and time, respectively.
We consider two system sizes, with number of particles $N=318^{2}$ and $512^{2}$, in a rectangular box with the side length ratio $L_{x}:L_{y}=2:\sqrt{3}$.
The density is fixed to $\rho=0.85$, and the temperature to $T = 0.5$. 

We equilibrate and sample the system in the canonical ensemble via molecular dynamics simulation. The equations of motion are integrated via a Verlet algorithm~\cite{Allen_book}, and the temperature is fixed via the Nos\'{e}-Hoover thermostat~\cite{Allen_book}. We perform the simulations with the GPU-accelerated GALAMOST package~\cite{Galamost}.

\begin{figure}[!!t]
 \centering
 \includegraphics[angle=0,width=0.45\textwidth]{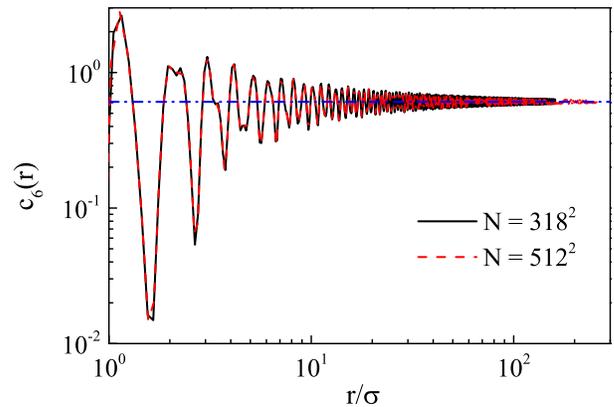}
 \caption{Bond-orientational correlation function $c_{6}(r)$ as a function of $r$ for $N=318^2$ (black solid line) and $N=512^2$ (red dashed line). The blue dash-dotted line marks the asymptotic value, $c_{6}(r \to \infty)=0.608$.
\label{fig:c6}
}
\end{figure}

To check for thermal equilibration, we compare the time evolution of runs started from a hexagonal lattice configuration, and from a random configuration, for the largest system we have considered, $N=512^{2}$.
In particular, we focus on the time evolution of the global bond-orientational order parameter $|{\bf \Psi}_{6}|=|\frac{1}{N}\sum_{j=1}^{N}{\bm{\psi}}_{6}({\bm r}_j)|$, where ${\bm \psi}_{6}(\bm{ r}_j)$ is the local bond-orientational order parameter of particle $j$ located at ${\bm r}_j$. 
This is defined as ${\bm \psi}_{6}({\bm r}_j)=\frac{1}{n}\sum_{m=1}^{n}\exp(i6\theta_{m}^{j})$, with $n$  the number of nearest neighbors of particle $j$, we determine via the Voronoi method, and $\theta_{m}^{j}$ is the angle between $({\bm r}_{m}-{\bm r}_{j})$ and a fixed arbitrary axis, we chose to be $\hat {\bm x}$.

Figure~\ref{fig:Equ} illustrates that, regardless of the initial configuration,  $|{\bf \Psi}_{6}|$ converges to $|{\bf \Psi}_{6}|\simeq0.780$ at $t\simeq 5\times10^{4}$, indicating that this time is enough for the system to reach thermal equilibrium.
Notice that equilibrium is reached in a much shorter time when the simulation starts from the hexagonal lattice, being the equilibrated state in the solid phase.
All data reported in the following are collected after a time $t=10^{5}$, ensuring thermal equilibration.

For the considered values of the control parameters, the system has been suggested to be in the solid phase~\cite{Hajibabaei2019}.
We explicitly show that this is the case investigating the bond-orientational correlation function $c_{6}(r=|\bm{r}_{i}-\bm{r}_{j}|)=\langle\bm{\psi_{6}}(\bm{r}_{i})\bm{\psi}_{6}^{*}({\bm r}_{j})\rangle$.
Fig~\ref{fig:c6} shows that $c_{6}(r)$ does not decay a 
large length scales, but converges to $0.608$, regardless of the system size. This is the expected behavior in the solid phase.
We also notice that $c_{6}(r)\simeq|{\bf \Psi}_{6}|^2$ at large $r$ (see Figs.~\ref{fig:Equ} and ~\ref{fig:c6}), indicating that $c_{6}(r)$ reaches its expected large $r$ limit.

\begin{figure}[tb]
 \centering
 \includegraphics[angle=0,width=0.45\textwidth]{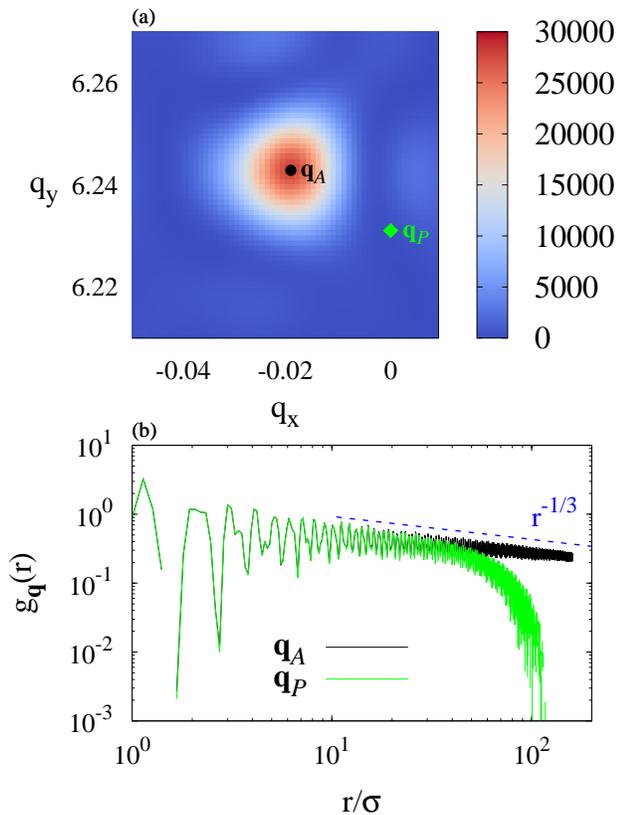}
 \caption{(a) Static structure factor around one of the six first Bragg peaks characterizing the solid phase. The green diamond indicates the wavevector $\bm{q}_P \simeq (0,6.231)$
at which a perfect hexagonal lattice at the same density would exhibit a peak.
 The black circle indicates the actual wavevector $\bm{q}_A \simeq (-0.01892, 6.24288)$ at which the peak occurs. 
 (b) translational correlation function $g_{\bm{q}}(r)$ evaluated at $\bm{q}_{A}$ (black) and at $\bm{q}_{P}$ (green), respectively. The data is for the system with $N=318^{2}$ particles. The dashed line is the KTHNY prediction for the decay of the translational correlation function in the solid phase.
\label{fig:Gqr}
}
\end{figure}

\section{Wavevector dependence \label{sec:wavevector}}

\begin{figure}[htb]
 \centering
 \includegraphics[angle=0,width=0.46\textwidth]{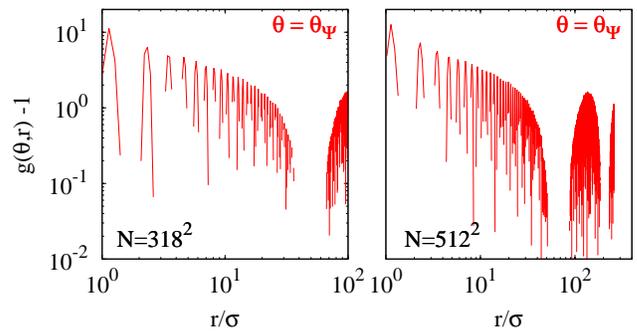}
 \caption{$g(\theta,r)-1$ as a function of $r$ for systems with $N=318^{2}$ (left column), and with $N=512^{2}$ (right column).
The angle $\theta$ is fixed to $\theta = \theta_\Psi$.
For this choice of $\theta$, the correlation function exhibits exponential-like decay even if the system is in the solid phase.
\label{fig:gpsi}
}
\end{figure}

The translational correlation function is defined as 
\begin{equation}
g_{\bf q}(r)=\frac{1}{2\pi r\Delta r\rho N}\sum_{j\neq k}\zeta(r-|{\bf r}_{j}-{\bf r}_{k}|)e^{i {\bf q}\cdot({\bf r}_{j}- {\bf r}_{k})}.
\label{eq:gqr}
\end{equation}
where $\zeta(r) = 1$ in the region $r \sim r+\Delta r$, $\rho$ is the number density, $r$ is the separation of a pair of particles, $\Delta r$ is the increment of $r$.
In numerical simulations, ${\bf q}$ is most often fixed to the wavevector ${\bf q}_P$ at which the structure factor of a perfect hexagonal lattice of density $\rho$, and orientation fixed by the simulation box, exhibits its main peaks ~\cite{subblock,Santi,weikaiqi,Wierschem2011,potential_softness}. 
In Fig.~\ref{fig:Gqr}(b) we show that, in the solid phase of the LJ system, this choice is not appropriate, as it leads to a translational correlation function which decays exponentially.

\begin{figure*}[htb]
 \centering
 \includegraphics[angle=0,width=0.9\textwidth]{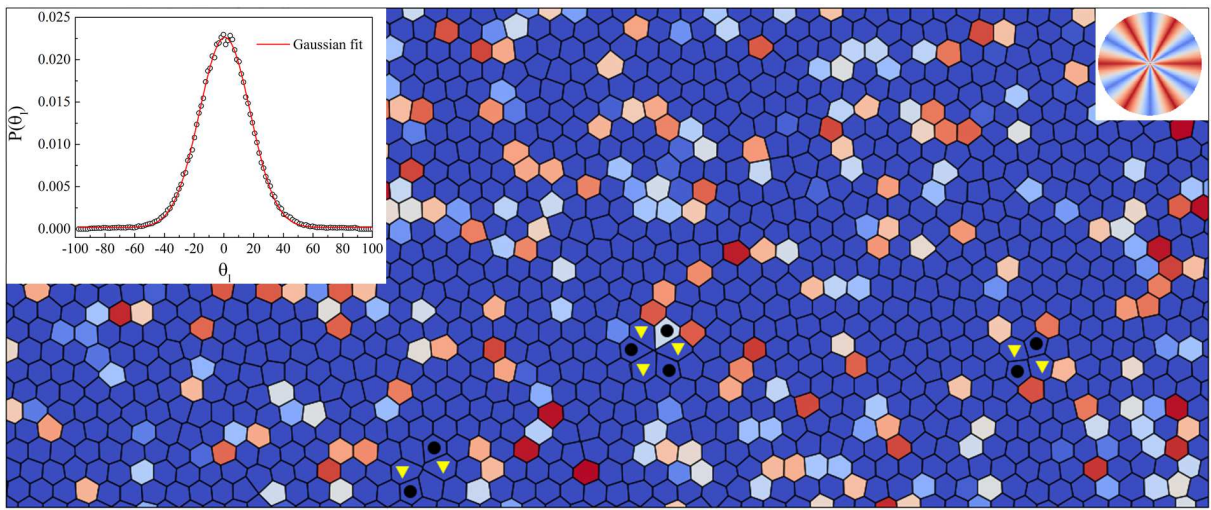}
\caption{Voronoi tessellation of a small fraction of a configuration of a 2d-LJ solid at $T=0.5$ and $\rho=0.85$. 
The color code reflects the direction of the bond-orientational order parameter associated to each particle, as indicated by the color wheel in the upper right inset.
Particles with 5 (7) neighbors are shown as black circles (yellow triangles). 
The upper left inset illustrates the distribution of the angle formed by the local bond-orientation parameter with the $x$-axis.
\label{fig:snap}
}
\end{figure*}

To rationalize this result, we evaluate the 2d static structure factor $S(q_{x}, q_{y})$, which is given by
\begin{equation}
S(q_{x}, q_{y})=\frac{1}{N}\langle\rho(q_{x}, q_{y})\rho(-q_{x}, -q_{y})\rangle,
\end{equation}
where $\rho(q_{x}, q_{y})$ is defined as
\begin{equation}
\rho(q_{x}, q_{y})=\sum_{j=1}^{N}\exp[i(q_{x}x_{j} + q_{y}y_{j})].
\end{equation}
Here, $x_{j}$ and $y_{j}$ correspond to $x$ and $y$ coordinates of particle $j$, respectively.
In Fig.~\ref{fig:Gqr}(a) we show $S(q_{x}, q_{y})$ for $N=318^{2}$, in a region around one of the six Bragg peaks. 
We do observe that the main peak occurs at a wavevector ${\bf q}_A$, which is shifted with respect to ${\bf q}_P$. An analogous results was previously found in hard disks~\cite{Krauth2011}. Here, however, we find a shift in both the magnitude and the orientation of the peak.
Figure~\ref{fig:Gqr}(b) shows that, when evaluated at the wavevector where the actual peak of the structure factor occurs, $g_{\bf q}(r)$ decays as a power-law.
The observed exponent is compatible with the KTNHY prediction, $\eta \simeq 1/3$.
Hence, to properly evaluate the degree of translational correlation through the investigation of $g_{\bf q}(r)$, care should be took in the selection of the proper wavevector. This can shift in both magnitude and orientation with respect to that of the ideal lattice. 

\section{Sample versus global-order orientation \label{sec:orientation}}
\begin{figure*}[tb]
\centering
 \includegraphics[angle=0,width=0.95\textwidth]{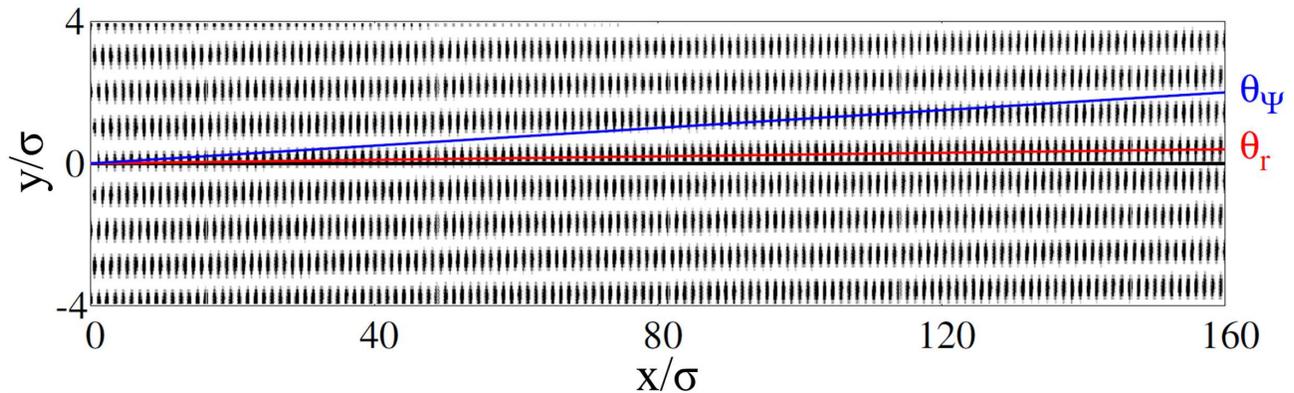}
 \caption{
 Two-dimensional pair correlation function $g(x, y)$ at $T=0.5$ and $\rho=0.85$. The black, red, and blue lines make an angle $\theta = 0,\theta_r$ and $\theta_\Psi$ with the $x$-axis, where $\theta_r$ identifies the symmetry of the crystal, and $\theta_\Psi$ corresponds to the direction of the global bond-orientational parameter. 
\label{fig:gxy}
}
\end{figure*}

The ${\bf q}$ dependence of above translational correlation function implies that, to average over different configurations, one first need to determine their structure factor, a somehow computational costly operation ($N \log N$).
To simplify this process, Bernard and Krauth~\cite{Krauth2011} introduced an alternative definition of the translational correlation function.
They suggested that, in the solid phase, crystals orient along the direction identified by the global bond-orientational order parameter, ${\bf \Psi}_{6}$, i.e. at an angle $\theta_\Psi$ from the ${ \hat{ \bf x}}$ axis, where ${\bf \Psi}_{6}\cdot {\hat{\bf x}} = |{\bf \Psi}_{6}| \cos(\theta_\Psi)$.
If this is so, then a suitable translational correlation function is given by a cut of the 2d correlation function along the ordering direction.
Formally, this is given by 
\begin{equation}
g(\theta, r)=\frac{1}{N}\sum_{j\neq k}\delta(r-(x_{j}-x_{k}))\delta(r\tan\theta-(y_{j}-y_{k})),
\end{equation}
with $\theta = \theta_\Psi$.
This method has been found robust in Ref.~\citenum{Krauth2011}, and it is appealing due to its simplicity, as one does not need to evaluate the Bragg peak of each configuration. It has indeed became very popular.

However, very recently Hajibabaei and Kim~\cite{Hajibabaei2019}, in numerical simulations of the same system we are considering here, found $g(\theta_\Psi,r)-1$ to exhibit exponential-like decay for configurations with long-range bond-orientational order, i.e. in the solid phase. 
We have found the same exponential decay, as illustrated in Fig.~\ref{fig:gpsi}.
We consider two different system sizes, $N=318^{2}$ (left column) and $N = 512^{2}$ (right column), to prove that this decay must not be attributed to the finite size of the considered system.
This result is in conflict with the predictions of KTHNY theory~\cite{KT, HN, Y}, and suggests that $g(\theta_\Psi,r)$ might not correctly track the degree of translational order of the system.

To rationalise the origin of this discrepancy, we illustrate in Fig.~\ref{fig:snap} a small fraction of the considered equilibrium sold-like configuration. 
The arrows indicate the orientation of the local bond-orientational parameter associated to each particle. It is visually clear that the system is in the solid phase, the directions of the local bond-orientational order parameters of distant particles being mostly parallel. In the figure, we also illustrate the topological defects, which are defined as particles that do not have $6$ neighbors as determined by Voronoi construction (see Fig.~\ref{fig:snap}). Interestingly, beside the commonly observed dislocation pairs (5-7-5-7 quartets) in e.g., hard systems~\cite{Glotzer,weikaiqi}, we also find more complex defects.

An apparent feature of Fig.~\ref{fig:snap}, possibly related to the existence of these complex defects, is the presence of large fluctuations in the orientation of the local bond-orientational parameter.
We quantify these fluctuations investigating the distribution of the local bond-orientational angles, $\theta_l$, with ${\bm \psi}_{6} ({\bf r}_j) \cdot \hat{ \bf x} = |{\bm \psi}_{6} ({\bf r}_j)|\cos( \theta_l^{(j)} )$.
The inset of Fig.~\ref{fig:snap} shows that the distribution is well described by a Gaussian function, with small deviations in the tails. 
We attribute to the large $\theta_l$ fluctuations also the  short-range oscillations of the bond-orientational correlation function, Fig.~\ref{fig:c6}, which are unusually pronounced.

The orientation of the global order parameter, $\theta_\Psi \simeq \ang{0.726}$, as well as the average local orientation, $\langle \theta_l\rangle \simeq \ang{0.750}$, will be affected by the fluctuations of the local bond-orientational order parameters, which conversely do not affect the orientation of the solid. Hence, one cannot expect $\theta_\Psi$ (or $\langle \theta_l\rangle$) to accurately identify the symmetry of the crystal. 
This might explain the exponential decay of the correlation function observed in Fig.~\ref{fig:gpsi}, which assumed the solid to be oriented along $\bf{\Psi}$.

\begin{figure}[tb]
 \centering
 \includegraphics[angle=0,width=0.45\textwidth]{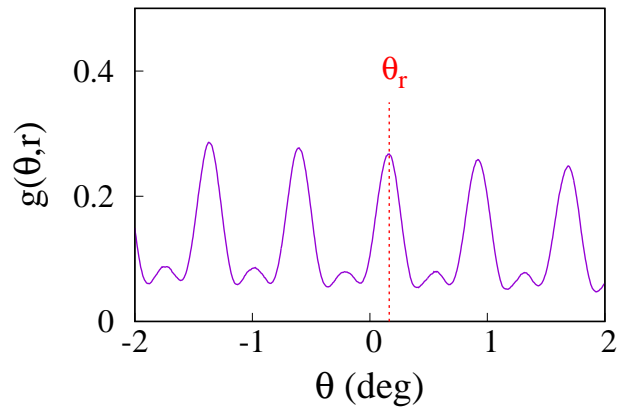}
 \caption{$\theta$ dependence of $g(\theta, r)$, for $r = 151.2\sigma$. The red dashed line marks the value of $\theta_{r}$. The data refer to a system with $N=318^{2}$ particles.
\label{fig:rdf2d}
}
\end{figure}

If neither $\theta_\Psi$ nor $\langle \theta_l\rangle$ identify the symmetry axis of the solid, then one need an alternative approach to determine it.
To this end, we investigate in Fig.~\ref{fig:gxy} the 2d pair correlation function $g(x, y)$, for $N=318^{2}$. 
Note that the figure is not in scale, and that we are focusing on a very narrow and long strip, of width $8$ and length $160$.
At short distances the peaks appear to lie on the $y=0$ line (black).
However, the figure clearly reveals that the axis of symmetry of the system is tilted by a small angle $\theta_{r}$ with respect to the $\hat x$ axis. 
A similar distortion is also apparent in Fig. S6 of Ref.~\citenum{Krauth2011}. 
To determine $\theta_r$, we study the $\theta$ dependence of the one-dimensional correlation function $g(\theta, r)$ of Eq.~\ref{eq:gqr}, at large $r$.
Figure~\ref{fig:rdf2d} shows that $g(\theta, r)$ peaks at different values $\theta$. The position of the first peak, $\theta_r \simeq \ang{0.158}$ in the figure, identifies the tilting angle of the crystal.

Figure~\ref{fig:gxy} visually confirms that the value of $\theta_r$ we have identified, rather than $\theta_\Psi$, corresponds to the tilting angle of the crystal.
We remark that a consistent estimate of the tilting angle is obtained comparing the peaks ${\bf q}_P$ and ${\bf q}_A$ of the static structure factor (see Fig.~\ref{fig:Gqr}),  ${\bf q}_P \cdot {\bf q}_A = |{\bf q}_P||{\bf q}_A| \cos(\theta_r)$.

\begin{figure}[tb]
 \centering
 \includegraphics[angle=0,width=0.46\textwidth]{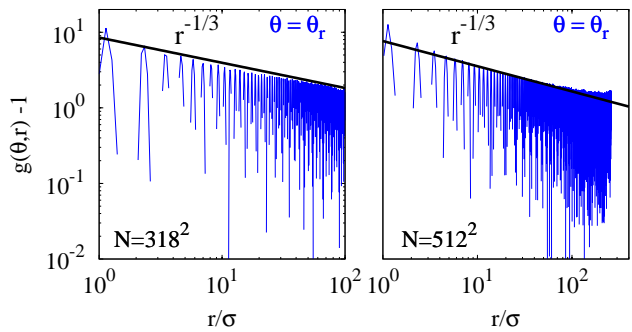}
 \caption{$g(\theta,r)-1$ as a function of $r$ for systems with $N=318^{2}$ (left column), and with $N=512^{2}$ (right column).
 The angle $\theta$ is fixed to $\theta_r$. 
 The full black lines are the KTHNY prediction for the decay of the translational correlation function in the solid phase.
\label{fig:gr}
}
\end{figure}

While the difference between $\theta_\Psi$ and $\theta_r$ is small, roughly half-degree in the case under investigation, this has important consequences for the behavior of the correlation functions.
We have indeed shown in Fig.~\ref{fig:gpsi} that $g(\theta_{\Psi},r)-1$ decays exponentially. 
Conversely, we show in Fig.~\ref{fig:gr} that $g(\theta_r,r)-1 \sim r^{-1/3}$.
This result is consistent with the KTHNY prediction, in the solid phase.

\section{Conclusions\label{sec:end}}
In summary, our results indicate that it is critical to correctly identify the orientation of 2d solids to correctly evaluate their translational correlation function. 
It is well known that 2d solids can be tilted with respect to symmetry axis one might expect given the boundary conditions, or equivalently given the shape of the simulation box~\cite{Wierschem2011}.
We have clarified that two approaches can be used to correctly identify the symmetry axis.
First, one might investigate the structure factor of the system, and infer the symmetry axis from the location of the fist peaks.
Secondly, one might perform a real space analysis, finding the angle of the first peak of the correlation function $g(r,\theta)$, at large $r$.

Importantly, our results clarify that the symmetry axis of the solid does not exactly coincide with the orientation of the global bond-orientational order parameter, as previously suggested~\cite{Krauth2011}. 
This is so as the global bond-orientation is generally affected by the stochastic fluctuations of the local bond-orientational order parameter. 
Our findings thus suggest that previous works have reported an exponentially decaying translational correlation function in the solid phase~\cite{Hajibabaei2019} as they have assumed the symmetry axis to be that fixed by the global bond-orientational order parameter.
Overall, our proposed analysis provides an approach to better estimate the location of the solid/hexatic transition.

We finally notice that, while we have certainly found the fluctuations of the local orientation to be important in 2d LJ solids, their actual relevance may be system specific. In particular, most previous works ~\cite{Krauth2011,Krauth2015,John_Russo,Glotzer} appear not be sensibly affected by the presence of these fluctuations. 
To rationalize this result, we notice that these studies focused on systems of particles interacting via purely repulsive forces. Hence, we speculate that the presence of attraction in the interparticle interaction may enhance the fluctuations of the local bond-orientational angle.

\begin{acknowledgments}
We acknowledge support from the Singapore Ministry of Education
through the Academic Research Fund (Tier 2) MOE2017-T2-1-066 (S) and from the
National Research Foundation Singapore, and are grateful to the National
Supercomputing Centre (NSCC) of Singapore for providing computational resources.
\end{acknowledgments}




\end{document}